# Tidal Dwarf Galaxies and missing baryons

Frederic Bournaud

CEA Saclay, DSM/IRFU/SAP, F-91191 Gif-Sur-Yvette Cedex, France

**Abstract** Tidal dwarf galaxies form during the interaction, collision or merger of massive spiral galaxies. They can resemble "normal" dwarf galaxies in terms of mass, size, and become dwarf satellites orbiting around their massive progenitor. They nevertheless keep some signatures from their origin, making them interesting targets for cosmological studies. In particular, they should be free from dark matter from a spheroidal halo. Flat rotation curves and high dynamical masses may then indicate the presence of an unseen component, and constrain the properties of the "missing baryons", known to exist but not directly observed. The number of dwarf galaxies in the Universe is another cosmological problem for which it is important to ascertain if tidal dwarf galaxies formed frequently at high redshift, when the merger rate was high, and many of them survived until today.

*In this article, I use "dark matter" to refer to the non-baryonic matter, mostly located in large dark halos – i.e., CDM in the standard paradigm. I use "missing baryons" or "dark baryons" to refer to the baryons known to exist but hardly observed at redshift zero, and are a baryonic dark component that is additional to "dark matter".*

## 1. Introduction: the formation of Tidal Dwarf Galaxies

A Tidal Dwarf Galaxy (TDG) is, per definition, a massive, gravitationally bound object of gas and stars, formed during a merger or distant tidal interaction between massive spiral galaxies, and is as massive as a dwarf galaxy [1] (Figure 1). It should also be relatively long-lived, so that it survives after the interaction, either orbiting around its massive progenitor or expelled to large distances. This requires a lifetime of at least 1Gyr, and a transient structure during a galaxy interaction would not deserve to be considered as a real TDG. The formation of TDGs in mergers has been postulated for decades [31], including potential candidates in the Antennae galaxies (NGC4038/39) [32], and became an increasingly active research topic after the study of these tidal dwarf candidates by Mirabel et al. [33].

Tidal tails are a common feature in galaxy interactions. There are also some tidal bridges and collisional rings that come about from the same processes and have similar properties, even though some details differ. The tidal tails are those long filaments seen around many interacting galaxies. They are made-up of material expelled from the disk of a parent spiral galaxy [2]. This material is expelled partly under the effect of tidal forces exerted by the other interacting galaxy, as the name suggests, but in fact for a large part by gravitational torques as the name does not suggest. The perturbing galaxy exerts non-radial forces in the disturbed disk, so that some gas loses angular momentum, flows towards the center where it can fuel a starburst [3], and some other gas gains angular momentum and flies away in long tidal tails.

At least two mechanisms can lead to the formation of massive substructures in tidal tails (Figures 1 and 2). First, gravitational instabilities can develop, as in any gas-rich medium. Having the material expelled from the disk eases gravitational collapse, as the stabilizing effect of rotation in the parent disk disappears, or weakens. The process is then a standard Jeans instability: when a region collects enough gas to overcome internal pressure support, it collapses into a self-bound object. The same process is believed to drive the formation of molecular clouds in spiral galaxy. The difference is that the interaction stirs and heats the gas, and increases its velocity dispersion [4]. As a result, the typical Jeans mass is high, enabling relatively massive objects to form. These star-forming gas clumps form at the Jeans mass, with a regular spacing, the Jeans wavelength, all along the long tidal tails. They resemble "beads on a string" [5]. Numerical simulations model their formation accurately, provided that the resolution is high (to revolve the instability length) and gas content is accounted for with some hydrodynamic model [6]. This mechanism can result in relatively numerous TDGs, maybe ten per major merger, but these are not very massive, at most a few $10^7 M_O$. The same process can form a large number of less massive structures, which are super star clusters rather than dwarf galaxies, potentially evolving into globular clusters.

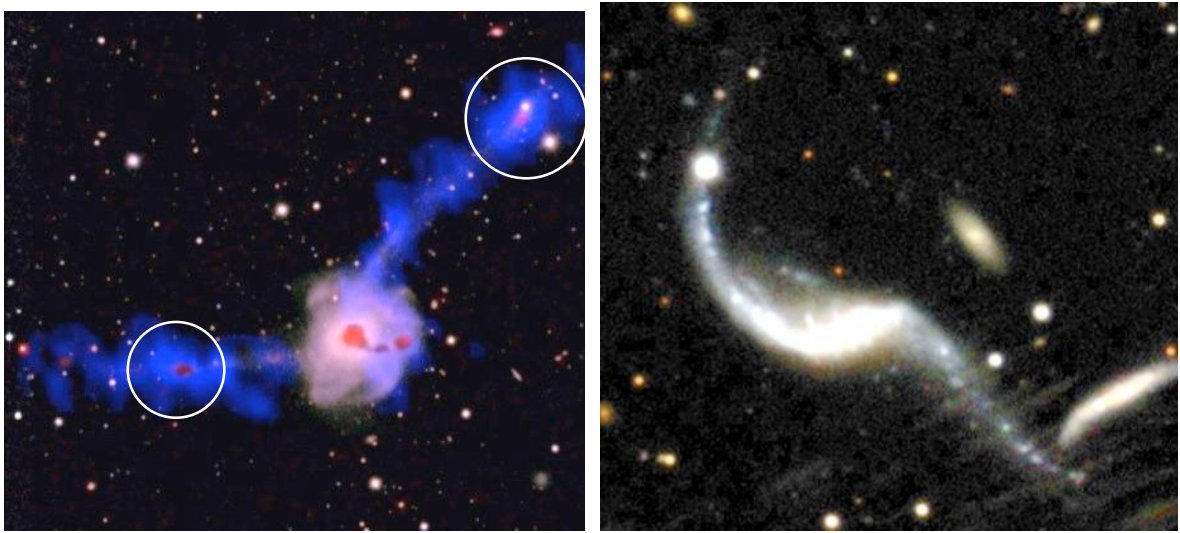

**Figure 1.** NGC7252 (left) is a recent merger of two spiral galaxies into a partially-relaxed central spheroidal galaxy. Two massive TDGs are found near the tip of the two long tidal tails (blue = HI, pink = Hα -- image courtesy of Pierre-Alain Duc). AM 1353-272 (right) does not have prominent, massive TDGs at the tip of tidal tails, but has instead may lower-mass objects all along its tail [37]. The bright spot on the northern tail is a foreground star.

Some tidal dwarfs are much more massive, a few $10^8$ or $10^9$ solar masses, and typically form as single objects at the tip of tidal tails [7]. They cannot form just by local instabilities in a tidal tail. These massive TDGs result from the displacement of a large region of the outer disk of the progenitor spiral galaxy into the outer regions of the tidal tail, where material piles up and remains or becomes self-bound [8,9]. As the interaction stirs the gas, the increased turbulence will provide the required pressure support to avoid fragmentation into many lower-mass objects. The shape of cold dark matter haloes, that are much more extended than the visible part of galaxies, make this process more efficient [10]. The pile-up of material in massive TDGs often occurs at the tip of tidal tails, but can also occur in different situations like gas captured and swung around the companion [34,5] or gas bridges linking two interacting galaxies [40,35].

A few tidal dwarfs of moderate mass, from the first mechanism, are found frequently in observed interacting galaxies. Massive TDGs in the very outer regions are more rare, at most 2-3 per merger, but some mergers do not have any at all. Observations suggest that the presence of one type of TDG reduces the number of TDGs of the other type [11]. This is likely because the formation of a massive, tip-of-tail TDG collects a large fraction of the gas, and less remains available for the formation of numerous low-mass TDGs along the tail, by the first mechanism [11].

Sections 2 and 3 will review the internal mass content of TDGs and how this can constrain the nature of missing baryons. Different constraints, arising from statistics on TDG formation and survival, will be discussed in Section 4.

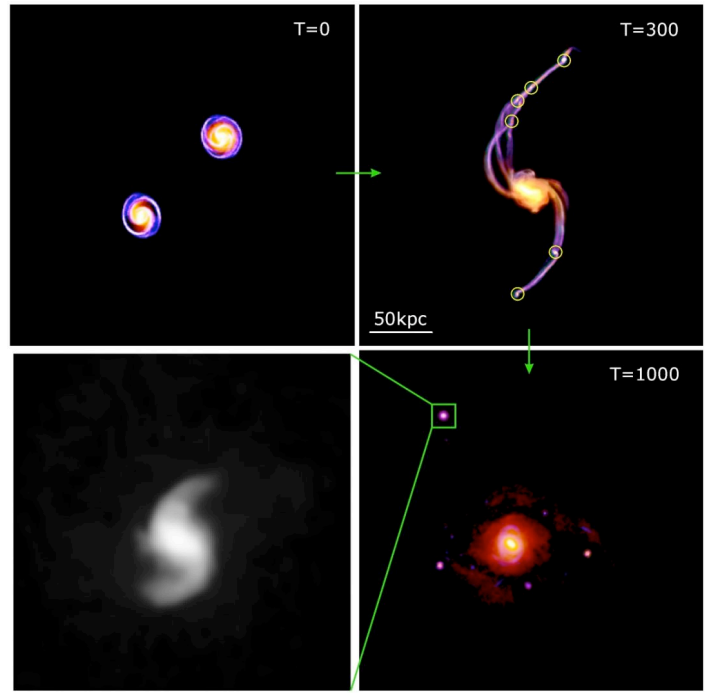

**Figure 2.** Simulation of the formation of long-lived TDGs, from the Bournaud & Duc sample [14]. The two colliding spiral galaxies are initially seen face-on. Yellow-red to blue colors code old stars versus young stars; time is in Myr. Massive TDGs form at the tip of the tails, and some lower-mass ones also form in particular in the Northern tail. A multi-grid technique increases the resolution on the most massive TDG to 10 parsec. It is resolved with a small internal spiral structure, and survives several Gyr around a massive red elliptical galaxy.

**2. Why should Tidal Dwarf Galaxies be free of dark matter?**

A common property of all TDGs, be they formed by a local instability, or by the gathering of a large portion of the progenitor spiral galaxy at the tip of a tidal tail, is that they are made-up only from material that comes from the *disk* of the parent spiral galaxy. Indeed, only the material initially in rotating disks, with velocity dispersions much lower than the rotation velocity, is strongly affected by tidal forces and gravity torques and forms tidal tails – and subsequently tidal dwarfs. Spheroids dominated by random velocity dispersions instead of rotation will barely develop a weak egg-shaped distortion during the interaction, but no long and dense tidal tail; this applies to the bulge of a spiral galaxy, a whole elliptical galaxy, but also to the dark matter halo of any galaxy.

Dark matter forms a spheroidal halo around galaxies, in both the standard Cold Dark Matter theories [12] and in other models (e.g., Warm Dark Matter [13]). All simulations show this dark matter cannot participate in the formation of TDGs [6,9,14]. Once a TDG has formed, its escape velocity is low, at most a few tens of km s$^{-1}$ for the biggest ones. The TDG will be embedded in the large halo of the parent spiral galaxy (since halos are much more extended than stellar and gaseous disks), so some dark matter particles will cross the TDG, but without being held by the gravitational well of the TDG. Indeed, the randomly-oriented velocities of dark matter particles in the halo of a spiral galaxy like the Milky Way are around 200 km s$^{-1}$, so the vast majority of these particles escape the tidal dwarf. At any instant, some dark matter particles from the halo will incidentally be at the position of the TDG, but this makes up only a negligible fraction of the mass of any TDG, at most a few percent.

TDGs should then be free of dark matter. This means that the dynamical mass measured from their rotation velocity and size, should not exceed their visible mass in stars and gas – in contrast with "normal" dwarf galaxies and spiral galaxies. Finding a dynamical mass in significant excess would mean that TDGs contain an unseen component. This could be dark matter only if (part of the) dark matter is in a rotating disk component in their progenitor spiral galaxies, just like stars and gas. Otherwise, this would require an unseen baryonic component in spiral galaxies, not just in the form of a hot gas halo – this one does not participate in the formation of TDGs – but in the form of very cold gas in the rotating disk. If not caused by some sort of matter, the dynamical mass excess could be attributed to Modified Gravity, which is a theoretical alternative to dark matter in all types of galaxies. These various possibilities will be discussed in the following section, in the context of observations of NGC5291.

## 3. The dynamical mass of Tidal Dwarf Galaxies

### 3.1 The collisional ring NGC5291

The collisional ring in NGC5291 was formed by a particular head-on collision, which formed a ring instead of the usual tails. Nevertheless, it formed Tidal Dwarfs just like more typical mergers with tails. The high metallicity of the dwarfs in this ring, and the young age of their stars, confirmed that they formed recently from tidal material. They are also too numerous to simply be "normal" dwarf galaxies that randomly happen to lie in the collisional ring (Figure 3).

The three largest TDGs in this system are spatially resolved in spectroscopic observations of their ionized Hα gas [7] and their neutral atomic (HI) gas [15]. The molecular gas observations by Braine et al. [36] did not require a dynamical mass larger than the visible mass. But new observations of the atomic gas [15] resolve internal velocity gradients in the TDGs, tracing their rotation up to their outermost regions. This confirms they are rotating, self-gravitating, decoupled from the large and diffuse HI ring in which they formed, and enables the dynamical mass to be measured from the rotation velocity and radius. This is what Bournaud et al. [15] did, and they found that the mass of these TDGs is in significant excess compared to their visible mass in stars, molecular gas, and atomic gas. The excess is just a factor of 2 or 3, not a factor 10 like in classical dwarf galaxies (Figure 3). This confirms the expected lack of dark matter in TDGs. Still, there should be an unseen component there, amounting to the mass of the visible one or even a bit more. Furthermore, rotation curves are surprisingly flat: the rotation velocity remains high far from the center of these three TDGs, while equilibrium with the visible mass can be achieved only if the velocity decreases in the outskirts. This adds evidence for the presence of some sort of unseen mass, mostly in the outer regions of these tidal dwarfs.

This observation remains compatible with most dark matter being in a large halo around spiral galaxies, but means that there was another dark component in the spiral galaxies from which the material now belonging to NGC5291's dwarfs was expelled.

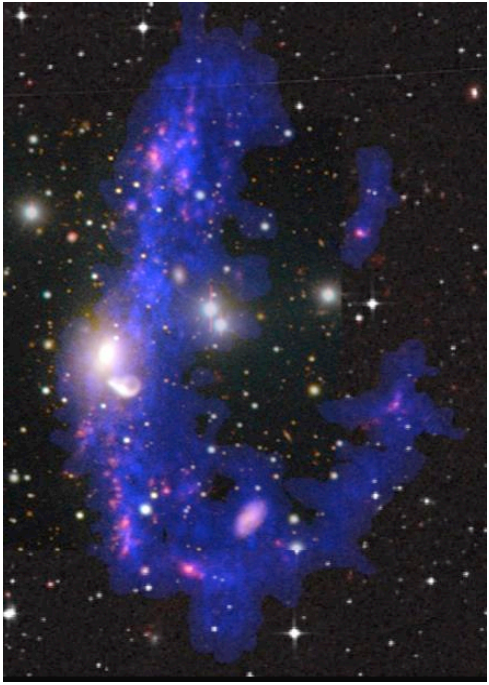
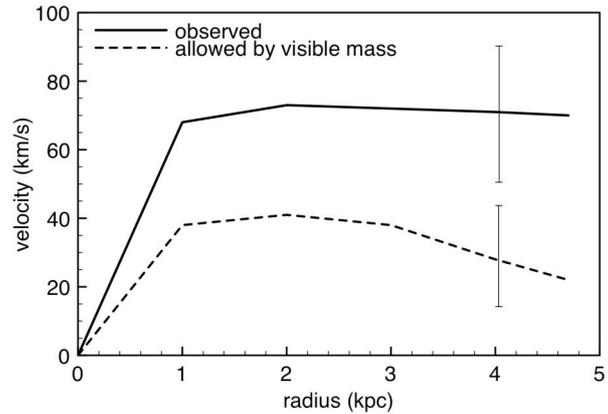

**Figure 3.** The collisional ring of NGC5291 and its tidal dwarfs (same as Fig.1) and the rotation curve of one of the TDGs [15]. The observed rotation velocity is too high, and the rotation curve too flat at large radius, to be accounted for by the visible mass distribution alone. Typical 1-sigma error bars are indicated. The rotation curve was measured from HI data and the visible mass estimated from stellar SEDs, HI and CO data.

*Unseen* molecular *gas?*

The "visible" mass of the TDGs in NGC5291 is for a large part atomic gas (HI), but also molecular gas. Most of the molecular gas mass is made-up of $H_2$, which cannot be directly observed. The molecular mass is traced by the emission of the CO molecule, using a standard "CO-to-$H_2$" conversion factor, which is somewhat uncertain. If the unseen mass in these TDGs is molecular gas, this would imply that the conversion factor changes much more than expected – by a factor 10, while the TDGs have a nearly-solar metallicity so no change by more than a factor of 2 or 3 was anticipated. A classical phase of molecular gas not well traced by CO would have other effects, like destabilizing the disk and triggering very active star formation, which is not observed [16]. More likely, the unseen component would be a non-standard phase of molecular gas, very cold and gathered in low-mass, dense but non-star forming $H_2$ blobs ([17,18], and Review by Pfenniger [38]). Finding it in TDGs would imply it was also present in the disk of the progenitor spiral galaxy, at least in its outer regions. This can be compatible with observations of the Milky Way [17,18,19] and the general dynamics of spiral galaxies [20], at least if one assumes this "dark molecular gas" comes in addition to a non-baryonic dark matter halo, not instead of it.

*Modified gravity?*

Another possible explanation to the large rotation velocities in NGC5291's TDGs is Modified Gravity (MOND). As it should affect any galaxy regardless of its origin (tidal or else), a high dynamical-to-visible mass ratio is naturally expected for TDGs in this context. While TDGs do not seem to have a dynamical-to-visible mass ratio as high as classical dwarfs, Gentile et al. [21] and Milgrom [22] have shown that the MOND theory can successfully account for their rotation curves.

*A disk of Cold Dark Matter?*

There could finally be another explanation that does not require a modification of the gravity, and not even any additional mass component. While dark matter is mostly in a spheroidal halo around spiral galaxies, there would be a part that is actually in a thick dark disk around the stellar disk. This could be brought in by small satellites that merge with the disk, and simulations by Read et al. [23] suggest such "dark disks" can be relatively massive (see also Purcell et al. [41]). In the CDM theory, this dark disk may naturally come in addition to the more massive spheroidal halo. Then, if the velocity dispersion of dark matter particles in this component is not too high, it could participate to the formation of TDGs just like the rest of the disk material (B. G. Elmegreen, priv. comm.). This hypothesis remains to be directly tested in simulations. It is in particular unclear how much of this disk dark matter will be dispersed during the merger, and how much will end-up in TDGs, potentially giving them a dark matter component massive enough to explain their observed rotation curves.

### 3.2 Other potential cases?

Error bars on NGC5291's data are such that the result is significant at ~2.5 sigma when one takes into account that 3 TDGs have been observed there. There are nevertheless errors that would affect the three observed TDGs in the same way, like the distance of the system, or its inclination. So the three "detections" are largely independent, but not completely. Can the result be confirmed in other systems?

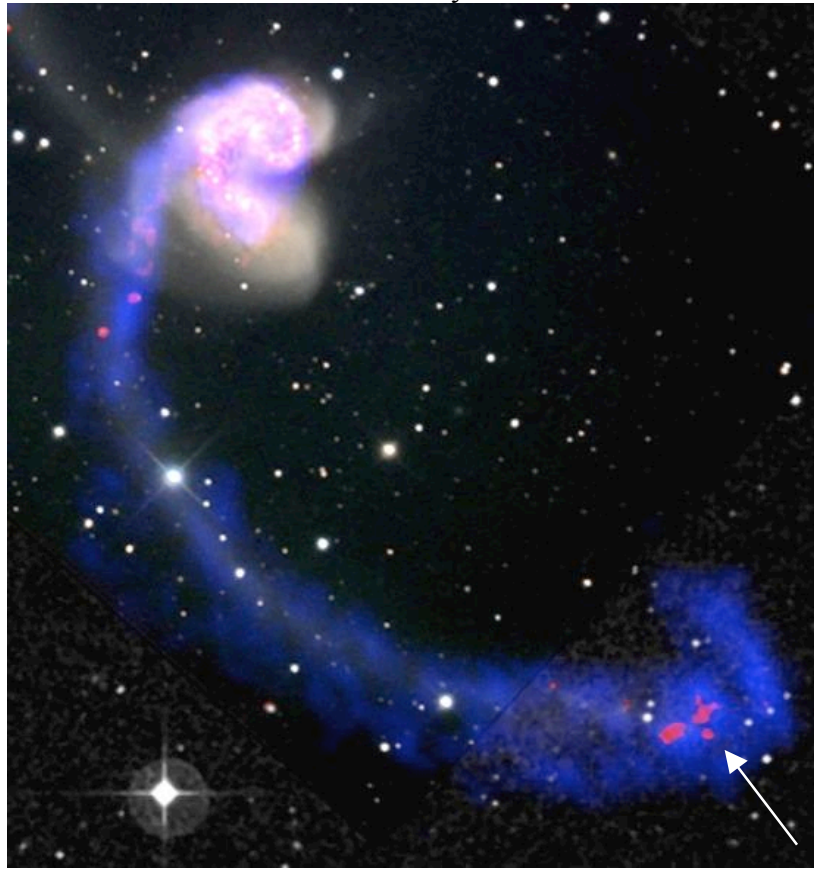

**Figure 4.** The Southern tail of the Antennae contains several star-forming clumps in a larger HI condensation. This, together with the almost face-on orientation of the system, makes it difficult to estimate a dynamical mass from the HI velocity curve. A recently revised distance estimate [24] may nevertheless indicate a dynamical mass higher than previously believed in the HI cloud. (Figure courtesy of Pierre-Alain Duc)

The most famous pair of colliding galaxies, the Antennae (NGC4038/39, Figure 4), has some tidal dwarf galaxies. It has long been thought that their dynamical and visible masses were similar. The original distance estimate to these galaxies could however be overestimated [24], which would make the dynamical mass two or three times higher than the visible mass, just as in NGC5291. The revision of the distance to the Antennae however remains largely uncertain [39]. There are other sources of uncertainty: first, TDGs in the Antennae are very young, still forming, maybe not at equilibrium. Rotation velocities may not trace the mass accurately – TDGs in NGC5291 were older, making the analysis more robust. Also, there is one blob of HI gas comprising three star forming regions. It is thus quite unclear if the HI velocities trace the total mass of these three objects, or the individual mass of each one – another problem that did not affect NGC5291, where each TDG corresponds to a single, resolved HI cloud. Another attempt in an old TDG in the Virgo Cluster [25] resulted again in a likely excess of dynamical mass, compared to the visible mass, but with large error bars.

Robust confirmations of an "unseen mass" in TDGs remain needed. They can be obtained, if relatively old TDGs are observed at high resolution and with a high signal-to-noise ratio, for instance with modern interferometers. This would definitely tell whether or not TDGs contain an unseen component, which has important implications, whatever the outcome, for the mass content of typical spiral galaxies.

## 4. Could (some) dwarf satellite galaxies be of tidal origin?

We show in the above section that the internal properties of TDGs (their visible mass compared to their rotation speed) may be related to the nature of some "missing" baryonic component. Tidal Dwarf Galaxies have another, completely different implication for the baryonic content of the Universe, which does not relate to their own internal properties, but to the number of dwarf galaxies formed by a tidal mechanism, compared to the total number of dwarf galaxies of any origin in the Universe.

TDGs may indeed contribute to the total population of dwarf galaxies, in particular dwarf satellites around massive galaxies – for instance some of the dwarf satellites of the Milky Way might, in principle, be tidal debris from collisions that occurred long ago. This would of course change the expected number of dwarf galaxies and the low-mass end of the mass function of galaxies – maybe not in the right direction if the predicted number of dwarf galaxies is already too high.

The key point is the survival of tidal dwarf galaxies. They must survive around, typically, one billion year for the merger in which they formed to be relaxed, so that they can appear as "normal" dwarf satellite galaxies around a "normal" galaxy (not an on-going merger). They must survive 5 or 10 Gyrs for mergers at high redshift to produce dwarf satellites at redshift zero. Is it frequently the case? Several factors can destroy TDGs: "internal" processes like the initial starbursts when gas-dominated TDGs begin to form stars and supernovae potentially eject their gas, and "external" processes, like a disruptive tidal field exerted by the very same galaxy that had formed the TDGs during an interaction.

Large samples with tens of simulations [14] are useful to tackle this question. The most massive tidal dwarf galaxies, that are large, massive ($10^{8-9}$ solar masses), rotating, formed preferentially at the tip of tidal tails, do not come in large numbers. Rarely more than

2 or 3 form in a major merger, sometimes none at all. Roughly half of them are destroyed within a couple of billions years, falling back onto their progenitor galaxy or being disrupted by its tidal field. For most of today's dwarf satellites to be TDGs from past mergers, one would need to form ~10 TDGs per major merger, each surviving a Hubble Time [26]. Hence, simulations suggest that only a modest fraction of modern dwarf satellite galaxies are of tidal origin – but not a completely negligible fraction: overall several percent. This fraction could be higher around red early-type galaxies, that experienced more mergers than spiral galaxies: this is because TDGs are expected to form mostly in mergers of spiral galaxies, and after the merger these progenitor galaxies generally become red early-type galaxies [42].

Lower mass TDGs that form with ~$10^6$ solar masses of baryons can be more numerous in each galaxy merger. They are more difficult to study in numerical simulations, as they require high spatial and mass resolution. Nevertheless, they can survive their initial starburst – but lose a large fraction of their mass [27]. They can also survive against the tidal field during more than a Gyr after their formation [28]. Simulations have never followed these low-mass objects for a long time; it seems plausible that some could form at high redshift and survive down to redshift zero, but probably as low mass remnants, potentially hard to detect at all, or evolving into compact, globular star clusters rather than dwarf galaxies [28].

While it then seems unlikely that the majority of today's dwarf galaxies are of tidal origin, they can have a significant contribution. There are known examples of TDGs forming at high redshift [29] and maybe some have survived down to redshift zero. Known examples of dwarf galaxies with unusual colors or metallicity could be long-lived tidal dwarfs [25,30]. The quest for robust and numerous cases of old tidal dwarfs remains open, however.

## 5. Summary

The formation of tidal dwarf galaxies (TDGs) is frequently observed in galaxy mergers, and has been extensively studied with the help of numerical simulations. While the long-term evolution and potential survival of these objects remains largely debated, their formation mechanisms are now well understood.

A clear prediction from all models is that TDGs cannot contain a significant mass fraction from the dark matter halo of their progenitor spiral galaxy. Observations suggest in several cases that the total, dynamical mass of TDGs exceeds their visible mass in gas and stars, which would indicate that they do contain some unseen component. This result, well established only for NGC5291, still needs a robust confirmation in other cases. An unseen component in TDGs could potentially constrain the presence of "missing baryons" in a cold gas phase.

Another cosmological implication of TDGs relates to their long-term survival, and how they can affect the mass function of dwarf galaxies. This question is still largely unsolved, but there is significant hope that modern numerical models, that can resolve the internal physics of TDGs in simulations of major mergers could lead to significant progress in the next years.

**Acknowledgements**
I acknowledge the editors of the "dwarf galaxies and cosmology" special volume for inviting me to write this tutorial review. Useful comments on an earlier version by Elias Brinks, Pierre-Alain Duc, Jonathan Braine, and two referees are appreciated, as well as discussions with Bruce Elmegreen, Moti Milgrom and Daniel Pfenniger on the origin of the mass discrepancy in tidal dwarf galaxies.

**References**

[1] Duc, P.-A. et al. 2000, AJ, 120, 1238

[2] Toomre, A. *in* Evolution of Galaxies and Stellar Populations, Beatrice M. Tinsley and Richard B. Larson Eds. New Haven: Yale University Observatory, 197, p.401

[3] di Matteo, P. et al. 2008, A&A, 492, 31

[4] Struck, C., Kaufman, M., Brinks, E., Thomasson, M., Elmegreen, B. G., Elmegreen, D. M. 2005, MNRAS, 364, 69

[5] Smith, B. J. et al. 2008, AJ, 135, 2406

[6] Wetzstein, M., Naab, T., Burkert, A. 2007, MNRAS, 375, 805

[7] Bournaud, F., Duc, P.-A., Amram, P., Combes, F., Gach, J.-L. 2004, A&A, 425, 813

[8] Elmegreen, B. G., Kaufman, M., Thomasson, M. 1993, ApJ, 412, 90

[9] Duc, P.-A., Bournaud, F., Masset, F. 2004, A&A, 427, 803

[10] Bournaud, F., Duc, P.-A., Masset, F. 2003, A&A, 411, L469

[11] Knierman, K. A. et al. 2003, AJ, 126, 1227

[12] Navarro, J. F., Frenk, C. S., White, S. D. M. 1996, ApJ, 462, 563

[13] Bullock, James S., Kravtsov, and Andrey V., Colín, Pedro 2002, ApJ, 564, L1

[14] Bournaud, F. & Duc, P.A. 2006, A&A, 456, 481

[15] Bournaud, F. et al. 2007, Science, 316, 1166

[16] Boquien, M. et al. A&A, 467, 93

[17] Pfenniger, D., Combes, F., Martinet, L. 1994, A&A, 285, 79

[18] Pfenniger, D., Combes, F. 1994, A&A, 285, 94

[19] Kalberla, P. M. W., Dedes, L., Kerp, J., Haud, U. 2007, A&A, 469, 511

[20] Revaz, Y., Pfenniger, D., Combes, F., Bournaud, F. 2009 A&A 501, 171

[21] Gentile, G. et al. 2007, A&A, 472, L25

[22] Milgrom, M. 2007, ApJ, 667, L45

[23] Read, J. I., Lake, G., Agertz, O., Debattista, V. P. 2008, MNRAS, 389, 1041

[24] Saviane, I., Momany, Y., da Costa, G. S., Rich, R. M., Hibbard, J. E. 2008, ApJ, 678, 179

[25] Duc, P.-A., Braine, J., Lisenfeld, U., Brinks, E., Boquien, M. 2007, A&A, 475, 187

[26] Okazaki, Tadashi, Taniguchi, Yoshiaki 2000, ApJ, 543, 149

[27] Recchi, S., Theis, C., Kroupa, P., Hensler, G. 2007, A&A, 470, L5

[28] Bournaud, F., Duc, P.-A., Emsellem, E. 2008, MNRAS, 389, L8

[29] Elmegreen, D. M., Elmegreen, B. G., Ferguson, T., Mullan, B. 2007, ApJ, 663, 734

[30] Michel-Dansac, L., Lambas, D. G., Alonso, M. S., Tissera, P. 2008, MNRAS, 386, L82

[31] Zwicky, F.: 1956, in: Ergebnisse der Exakten Naturwissenschaften 29, 344

[32] Schweizer, F.: 1978, in Structure and Properties of Nearby Galaxies, IAU Symp. 77, E.M. Berkhuijsen and R. Wielebinski Eds., p. 279.

[33] Mirabel, I. F., Dottori, H., Lutz, D. 1992, A&A, 256, L19

[34] Bournaud, F., Combes, F. 2003, A&A, 401, 817

[35] Hancock, M., Smith, B. J., Struck, C., Giroux, M. L., Hurlock, S. 2009, AJ, 137, 4643

[36] Braine, J., Duc, P.-A., Lisenfeld, U., Charmandaris, V., Vallejo, O., Leon, S., Brinks, E. 2001, A&A, 378, 51

[37] Weilbacher, Peter M., Fritze-v. Alvensleben, Uta, Duc, Pierre-Alain, Fricke, Klaus J. 2002, ApJ, 579, L79

[38] Pfenniger, D. 2004, in IAU Symp. 220, S. D. Ryder, D. J. Pisano, M. A. Walker, and K. C. Freeman Eds., p.241

[39] Schweizer, F. et al. 2008, AJ, 136, 1482

[40] Koribalski, B. , & Dickey, J. M. 2004, MNRAS, 348, 1255

[41] Purcell, C. W., Bullock, J. S., Kaplinghat, M. 2009, ApJ, submitted. arXiv:0906.5348

[42] Martig, M., Bournaud, F., Teyssier, R., Dekel, A. 2009, ApJ submitted. arXiv:0905.4669